\documentstyle[12pt]{article}

\def\NPB#1#2#3{Nucl. Phys. B{#1} (19#2) #3}
\def\PLB#1#2#3{Phys. Lett. B{#1} (19#2) #3}

\def\PRL#1#2#3{Phys. Rev. Lett. {#1} (19#2) #3}
\def\PRT#1#2#3{Phys. Rep. {#1} C (19#2) #3}
\def\MODA#1#2#3{Mod. Phys. Lett.  {#1} (19#2) #3}
\def\yzero{\smash{\hbox{$y\kern-4pt\raise1pt\hbox{${}^\circ$}$}}}
\def\-{\hphantom{-}}
\def\ov{\overline}
\def\beq{\begin{equation}}
\def\eeq{\end{equation}}
\def\beqa{\begin{eqnarray}}
\def\eeqa{\end{eqnarray}}

\def\IF{\relax{\rm I\kern-.18em F}}
\def\II{\relax{\rm I\kern-.18em I}}
\def\IP{\relax{\rm I\kern-.18em P}}

\def\NN{{\cal N}}

\def\IR{\bf R}

\def\IZ{\bf Z}
\def\IS{\bf S}
\def\IT{\bf T}
\def\RP{\bf RP}
\def\AdS{\bf AdS}

\advance\textheight by \topskip
\begin{document}

\makeatletter
\@addtoreset{equation}{section}
\makeatother
\renewcommand{\theequation}{\thesection.\arabic{equation}}
\pagestyle{empty}
\rightline{CERN-TH/99-397}

\rightline{\tt hep-th/9912145}
\vspace{1.5cm}
\begin{center}
\LARGE{Comments on Non-supersymmetric Orientifolds \\
at Strong Coupling \\[10mm]}
\large{
Angel~M.~Uranga \footnote{\tt angel.uranga@cern.ch}\\[2mm]}
{\em Theory Division, CERN}\\
{\em CH-1211 Geneva 23, Switzerland} \\[4mm]

\vspace*{2cm}

\small{\bf Abstract} \\[7mm]
\end{center}

\begin{center} \begin{minipage}[h]{14.0cm}
{\small
We consider several properties of a set of anti-D$p$-branes in the
presence of orientifold $p$-planes in type II theory. This system breaks
all the supersymmetries of the theory, but is free of tachyons. In
particular, we center on the case of a single anti-D$p$-brane stuck at a
negatively charged orientifold $p$-plane, and study its strong coupling
behaviour for $p=2,3,4$. Interestingly enough, as the coupling
increases the system undergoes a phase transition where an additional
antibrane is created. We conclude with some remarks on the limit of large
number of antibranes on top of orientifold planes.
}

\end{minipage}
\end{center}

\newpage
\setcounter{page}{1}
\pagestyle{plain}
\renewcommand{\thefootnote}{\arabic{footnote}}
\setcounter{footnote}{0}

\section{Introduction}

In this paper we will be interested in the properties of a set of
anti-D$p$-branes (denoted $\ov{Dp}$-branes) in the neighbourhood of an
orientifold $p$-plane (O$p$-plane) in type II string theory (see
\cite{orient} for early discussions on orientifolds, and
\cite{tasi} for a review). Since the antibranes and the orientifold
projection preserve different sets of supersymmetries, the system breaks
all the supersymmetries of the theory, but it is free of tachyons.

The motivation to study these systems is two-fold. First, even though they
are non-supersymmetric, they are relatively simple. For instance,
supersymmetry is preserved on the closed string sector, and bulk physics
reduces to that of type II theory. These systems may therefore be a good
laboratory to continue extending our limited understanding of string
theory and string duality in non-supersymmetric situations. In fact, we
will be able to extract information about the strong coupling behaviour of
these systems in particular cases. The second motivation is that
configurations with orientifold planes and antibranes appear in the
non-supersymmetric (but tachyon-free) type I compactifications in
\cite{nonsusy} (see also \cite{sugimoto}). Models of this kind exhibit
certain phenomenologically interesting features, and deserve further study. 
Our comments in the present paper constitute a small step towards dealing
with some of the relevant issues in a simpler and more controlled situation.

This note is organized as follows. In Section~2 we make some remarks on
the perturbative properties of these configurations, and compute the 
leading contribution to the interaction between antibranes and the 
orientifold plane, which is relevant for the stability of the configurations. 
In Sections 3 we study duality properties of these models, and in
particular the strong coupling behaviour of a single $\ov{Dp}$-brane stuck 
on top of a negatively charged O$p$-plane, for $p=2,3,4$. This is the
simplest non-supersymmetric orientifold configuration within our
framework. Using dual descriptions we show that at strong coupling an
additional antibrane is created. Section~4 contains some final comments.

\section{Weak coupling description}

Recall the configuration of $N$ D$p$-branes on top of an O$p$-plane, which
preserves sixteen supersymmetries. There are two kinds of orientifold
projections in string perturbation theory, which differ in the sign of the
contribution of the $\RP_2$ world-sheet topology, and hence in the RR
charge of the corresponding O$p$-planes. We denote by O$p^{\pm}$-plane the
orientifold plane with $\pm 2^{p-4}$ units of D$p$-brane charge (as counted 
in the covering space) \footnote{As further discussed in Section~3, the
O$p^+$-plane usually comes in two varieties, distinguished by the value of
a RR
flux. Since they are identical in perturbation theory, we will not 
distinguish them in the present section.}. The massless open string modes
produce a world-volume gauge group $G$, along with the scalars and fermions 
required to fill a vector multiplet of the corresponding supersymmetry. The 
group $G$ is $SO(N)$ or $USp(N)$ for the case of O$p^-$-plane or 
O$p^+$-plane, respectively.

Let us consider instead a set of $N$ $\ov{Dp}$-branes on top of the
O$p$-plane. Before the orientifold projection, the massless spectrum on
the $\ov{Dp}$-brane world-volume consists of a $U(N)$ vector multiplet with
respect to the sixteen unbroken supersymmetries. As discussed in
\cite{sugimoto}, the orientifold projection on the bosonic fields is just
as for D$p$-branes, while fermions pick up an additional minus sign
\footnote{This sign is related, by open-closed duality, to the fact that
antibranes and branes carry opposite RR charges.}. The world-volume
massless fields are given in the following table
\begin{center}
\begin{tabular}{|l||c|c|c|}
\hline
             & $SO(p-1)$ & $SO(9-p)$ & $SO(N)\, $ ; $USp(N)\, $ \\
\hline\hline
Gauge bosons & vector & singlet & $N(N-1)/2$ ; $N(N+1)/2$\\
\hline
Scalars & singlet & vector &  $N(N-1)2$ ; $N(N+1)2$ \\
\hline
Fermions & spinor & spinor &  $N(N+1)2$ ; $N(N-1)2$ \\
\hline
\end{tabular}
\end{center}
The quantum numbers are with respect to the $SO(p-1)$ Lorentz little
group, a $SO(9-p)$ global symmetry (arising from rotational invariance in
the transverse space), and $SO(N)$ or $USp(N)$ gauge group for O$p^-$- or
O$p^+$-planes, respectively (Notice that the symmetric representation of
$SO(N)$ and the antisymmetric of $USp(N)$ are actually reducible). As is
manifest from the spectrum, the system breaks all the supersymmetries. 
However, and in contrast with the more familiar brane-antibrane 
configurations, the spectrum contains no tachyons, since no annihilation
can take place. 

Due to lack of supersymmetry, the flat directions  of the scalar potential
are not protected against further corrections, which therefore control the
stability of the configuration. At leading order, they arise from the
M\"obius strip, which is the simplest world-sheet topology feeling the
breaking of all supersymmetries. The corresponding piece in the partition
function is related to the interaction energy between the O$p$-plane and
the $\ov{Dp}$-branes. The answer expected from long-distance
considerations (oppositely charged objects attract and equally charged
objects repel) turns out to be correct even at short distance, as we
sketch in the following.

Consider for simplicity a single $\ov{Dp}$-brane (and its image) located
at a the position $\vec{X}$ in the $(9-p)$-dimensional transverse space. For 
a D$p$-brane the M\"obius strip contribution would be given by (see
\cite{tasi} for conventions)
\beqa
{\cal A}_{\cal M} & = & \pm V_{p+1} \int_{0}^{\infty} \frac{dt}{2t}
(8\pi^2 \alpha' t)^{-\frac{p+1}{2}} e^{-\frac{2X^2 t}{\pi \alpha'}}
q^{-\frac 23} \prod_{n=1}^{\infty}(1-q^{2n} e^{-i\pi n})^{-8} \times
\nonumber \\
& & \frac 12 \left\{ -iq^{-\frac 13} \prod_{n=1}^{\infty} (1+q^{2n-1} 
e^{-i\pi(n-1/2)})^8 + i q^{-\frac 13} \prod_{n=1}^{\infty} (1-q^{2n-1} 
e^{-i\pi(n-1/2)})^8 + \right. \nonumber\\
& & \;\; \left. + 16 q^{\frac 23} \prod_{n=1}^{\infty} (1+q^{2n}
e^{-i\pi n})^8 \right\} 
\label{susymob}
\eeqa
where $q=e^{-\pi t}$ and the $\pm$ sign corresponds to the O$p^{\mp}$-plane 
case. Due to supersymmetry, the first two contributions in the bracket,
arising from the NS sector, cancel the remaining one, from the R sector.
The amplitude in the case of $\ov{Dp}$-branes differs just in the sign of
the R sector contribution (due to the additional sign in the $\Omega$ 
action on spacetime fermions). Therefore, it is given by minus two times
the above R contribution, and can be written as
\beqa
{\cal A}_{\cal M} & = & \mp V_{p+1} \int_{0}^{\infty} \frac{dt}{2t}
(8\pi \alpha' t)^{-\frac{p+1}{2}}\, e^{-\frac{2X^2 t}{\pi \alpha'}} \,
F(q^2)
\label{nonsusymob}
\eeqa
with $F(q^2)=\frac{f_2(q^2) f_4(q^2)}{f_1(q^2) f_3(q^2)}$, and the
functions $f_i(x)$ defined as in \cite{tasi}. For non-zero $X$, we can
change variables to get
\beqa
{\cal A}_{\cal M} & = & \mp \frac 12 V_{p+1} (8\pi \alpha't)^{-\frac{p+1}{2}}
(\frac{\pi \alpha'}{2X^2})^{-\frac{p+1}{2}} \int_{0}^{\infty} du \,
u^{-\frac{p+3}{2}}\, e^{-u} \, F(e^{-\frac{\pi^2 \alpha' u}{X^2}}) 
\label{final}
\eeqa
This integral converges for $-1 < p < 7$, as follows from the asymptotic
behaviour
\beqa
F(e^{-2\pi t}) & {\stackrel{t\to\infty}{\longrightarrow}} & 16 \nonumber\\
F(e^{-2\pi t}) & {\stackrel{t\to 0 }{\longrightarrow}} & 256\, t^4
\eeqa
At large $X$, keeping only the leading term in $F$, the amplitude reads
\beqa
{\cal A}_{\cal M} = \mp 2^{p-4} V_{p+1}\, 2\pi (4\pi^2\alpha')^{3-p}\, 
G_{9-p}(X^2)
\label{long}
\eeqa
where $G_{9-p}(X^2)=\frac 14 \pi^{\frac{p-9}{2}}\Gamma(\frac{7-p}{2}) 
|X|^{p-7}$ is the $(9-p)$-dimensional massless scalar Green's function.
The force between the objects goes like $d{\cal A}_{\cal M}/dx$, hence the
amplitude (\ref{long}) corresponds to a repulsive (resp. attractive)
interaction between $\ov{Dp}$-branes and O$p^-$-planes (resp. 
O$p^+$-planes), due to exchange of massless closed string modes in the
transverse $9-p$ directions. Comparing (\ref{long}) with the brane-brane
interactions in \cite{tasi}, the additional factor of $\mp 2^{p-4}$
accounts for the orientifold charge and tension.

At small values of $X$, replacing $F$ in (\ref{final}) by its leading
term does not give a good approximation to the complete integral. A
better picture of the interaction is obtained by expanding the original
expression (\ref{nonsusymob}) around $X=0$, 
\beqa
{\cal A}_{\cal M} & = & \mp \left[ \Lambda - M X^2 + O(X^4) \right]
\label{short}
\eeqa
with positive coefficients
\beqa
\Lambda & = & V_{p+1} \int_0^\infty \frac{dt}{2t}
(8\pi\alpha')^{-\frac{p+1}{2}} F(q^2) \nonumber \\
M & = & V_{p+1}\frac{4}{\pi\alpha'} \int_0^\infty \frac{dt}{2}
(8\pi\alpha')^{-\frac{p+1}{2}} F(q^2) 
\eeqa
We see that also at short distances the interaction is repulsive (vs.
attractive) for the O$p^-$-plane (O$p^+$-plane) case. Notice that the
$X^2$ contribution can be interpreted in the open string channel as a
one-loop correction to the mass of the scalar $X$, which was massless at
(open-string) tree level.

From our above comments, we learn that the configuration of $N$
$\ov{Dp}$-branes on top of an O$p^+$-plane is stable at this order, and so
at sufficiently small coupling. On the other hand, the configuration of
$N$ $\ov{Dp}$-branes on top of an O$p^-$-plane is unstable, with the
exception of the case $N=1$, where the brane is stuck on the orientifold
even at tree level. One might worry about the consistency of the latter
configuration, since it involves coincident charges of the same kind.
However, at short distances the interaction arises from (\ref{short})
rather than from the Coulomb-like (\ref{long}), and is finite for $X=0$. We 
would like to stress that the absence of short-distance divergences
follows from the fact that the model contains no  open string tachyons, in
contrast with brane-antibrane systems, which are singular in that regime
\cite{banks}. 

\section{Strong coupling behaviour}

In this Section we consider the strong coupling behaviour of the
configurations of antibranes on top of orientifold planes. For obvious
reasons we will be more interested in configurations which are at least
perturbatively stable, and in particular we will center on the system of a
single $\ov{Dp}$-brane stuck on an O$p^-$-plane.

We will base our arguments on string duality, which has been a useful tool
in analyzing the strong coupling behaviour of supersymmetric configurations
of D$p$-branes and O$p$-planes. The case of O3-planes has been discussed
in \cite{witthree} using type IIB self-duality (see also \cite{egkt}), the
M-theory lifts of O4-planes have been determined in \cite{hori} (see also
\cite{gimon}), and those of O2-planes and O0-planes have been considered
in \cite{sethi} and \cite{hanany}, respectively. Useful information about
other values of $p$ can be extracted from \cite{witfive} for O5-planes,
\cite{landlop} for O6-planes and \cite{witseven} for O7-planes. We expect
these results to help in understanding duality properties in our
non-supersymmetric models, since the bulk is still supersymmetric, and its
duality properties may extend to the fixed points of the orientifold
action \footnote{In other words, the orientifolding action in our models
belongs to family 2 in the classification in \cite{sen}, where it was 
argued that the quotient theory retains the duality properties of the
original theory, {\em i.e.} `orientifolding commutes with duality'.}. In
the following sections and for illustrative purposes, we center on the
particular case of O3-, O4- and O2-planes.

\subsection{Orientifold 3-planes}

It will be useful to recall the situation for supersymmetric configurations 
of D3-branes on O3-planes, studied in \cite{witthree}. There are four
types of supersymmetric configurations, labeled by
$(\theta_{NS},\theta_{R})$, where $\theta_{NS},\theta_{R}=0,\frac 12$
denote the field-strength flux of the type IIB 2-forms in the transverse
space (with the origin excised) $\RP_5\times \IR$. The map between the
configurations and their fluxes is
\begin{center}
\begin{tabular}{|l|c|c|c|}
\hline
D-brane description & $(\theta_{NS},\theta_R)$ & RR charge & World-volume\\
\hline\hline
O3$^-$ + 2P D3     & $(0,0)$ & $2P-1/2$ & $SO(2P)$ \\
\hline
O3$^-$ + (2P+1) D3  & $(0,1/2)$ & $2P+1/2$ & $SO(2P+1)$ \\
\hline
O3$^+$ + 2P D3     & $(1/2,0)$ & $2P+1/2$ & $USp(2P)$ \\
\hline
${\widetilde{O3}}^+$ + 2P D3 & $(1/2,1/2)$ & $2P+1/2$ &
$USp(2P)$ \\
\hline 
\end{tabular}
\end{center}
where the ${\widetilde{O3}}^+$-plane is an exotic variety of the
$O3^+$-plane, differing from it in a RR-flux, and producing also a
$USp(2P)$ gauge theory. The above configurations come in multiplets of the
type IIB $SL(2,\IZ)$ duality group. The $SL(2,\IZ)$ action on the
configurations follows from its action on the corresponding NS-NS and R-R
fluxes,
and underlies the Montonen-Olive duality properties of the $\NN=4$
supersymmetric gauge theories in the last column. Setting $P=0$,
$SL(2,\IZ)$ also gives information about the non-perturbative properties
of O3-planes. For instance, their behaviour at strong coupling can be
extracted from their duals under the $\tau\to -1/\tau$ transformation. We
thus learn that in the strong coupling limit the O3$^-$-plane and the
${\widetilde{O3}}^+$-plane are unchanged, whereas the O3$^+$-plane turns
into an O3$^-$-plane with a stuck D3-brane, and vice-versa (as proposed
earlier in \cite{egkt}).

\medskip

Let us now turn to the non-supersymmetric case of $\ov{D3}$-branes on
O3-planes. The classification of these configurations is analogous to that
in the supersymmetric case, for the following reason. In the
non-supersymmetric configurations, the transverse space (after excising
the origin) is also $\RP_5\times \IR$. Moreover, as mentioned in Section~2, 
the non-supersymmetric theories differ from the sypersymmetric ones only in
an additional minus sign in the orientifold action on fermions. This means
that fermions pick up an additional minus sign in going along 
non-contractible 1-cycles in $\RP_5$, but the bosonic properties of the
background are unchanged at the classical level, and so is the
classification of fluxes for the 3-form fields stregths. Finally, since
these fluxes are discrete, topological, this classification cannot be
changed by quantum corrections, even in the non-supersymmetric situation.
Therefore, we obtain four types of non-supersymmetric configurations, as
follows
\begin{center}
\begin{tabular}{|l|c|c|c|}
\hline
D-brane description & $(\theta_{NS},\theta_R)$ & RR charge & World-volume\\
\hline\hline 
O3$^-$ + $2P \, {\ov{D3}}$     & $(0,0)$ & $-2P-1/2$ & $SO(2P)$ \\
\hline
O3$^-$ + $(2P+1)\, {\ov{D3}}$  & $(0,1/2)$ & $-2P-3/2$ &
$SO(2P+1)$\\
\hline
O3$^+$ + $(2P+2) \, {\ov{D3}}$   & $(1/2,0)$ & $-2P-3/2$ &
$USp(2P+2)$\\
\hline
${\widetilde{O3}}^+$ + $(2P+2) {\ov{D3}}$ & $(1/2,1/2)$ &
$-2P-3/2$ &$USp(2P+2)$ \\
\hline
\end{tabular}
\end{center}
Recall that scalars transform in the adjoint of the gauge group, but fermions 
do not. As in the supersymmetric case, these configurations must appear in
$SL(2,\IZ)$ multiplets, and therefore transform according to their flux
structure. Notice that in the above table we have arranged the number of
$\ov{D3}$-branes so that configurations in the same $SL(2,\IZ)$ multiplet
have the same RR charge.

This leads to interesting proposals for the strong coupling behaviour of
the configurations. For instance, a configuration of $N$ $\ov{D3}$-branes
on a O3$^+$-plane, which is stable at weak coupling, becomes unstable at
sufficiently strong coupling, since it is better described as a set of
$N+1$ $\ov{D3}$-branes on a O3$^-$-plane at weak string coupling. To stay
on the safe side, in the following we center on a particular case which is
stable at weak and strong coupling (in the hope that it behaves nicely
also in between), namely the configuration of an O3$^-$-plane with a stuck
$\ov{D3}$-brane. This object has charge $-\frac 32$ under the RR four-form, 
and corresponds to fluxes $(0,\frac 12)$. We propose that at strong coupling 
this configuration turns into a set of two $\ov{D3}$-branes on an
O3$^+$-plane, which has the appropriate charge and flux structure. Notice
that in the latter configuration, the $\ov{D3}$-branes are bound to the
O3$^+$-plane due to the attractive interactions discussed in Section~2,
and only in the extreme strong coupling they are free to move off into the
bulk (the dual coupling being strictly zero in this case). Hence this
model presents an interesting transition between two mechanisms to bind
antibranes to orientifold planes (stuck antibranes vs. attracted
antibranes).

A further bit of information supporting this proposal follows from the
world-volume perspective. Even though the dynamics of the relevant field
theory is non-supersymmetric and therefore intractable beyond weak
coupling, certain quantities, namely anomalies of global symmetries,
should match in the weak and strong coupling limit \cite{thooft}. In the
present case, there is a classical $SU(4)$ symmetry associated to
rotations in the six-transverse dimensions. In the configuration of an
O3$^-$-plane with an stuck $\ov{D3}$-brane, the world-volume contains no
bosonic fields, but there is a fermion transforming in the fundamental
representation of this $SU(4)$, and leading to an anomaly. In the
configuration of an O3$^+$-plane with two $\ov{D3}$-branes, there is a
gauge group $USp(2)$, under which scalars transform in the adjoint, but
under which fermions are singlets. The latter transform in the fundamental
of the $SU(4)$ global symmetry. Hence the anomalies for both
configurations match, making our strong coupling proposal plausible.

An intuitive explanation for the creation of an additional $\ov{D3}$-brane 
would be as follows. We start with one $\ov{D3}$-brane stuck at an
O3$^-$-plane. As the string coupling becomes stronger, it becomes easier
to nucleate D3-$\ov{D3}$-brane pairs out of the vacuum. Since the `real'
and the `virtual' $\ov{D3}$-branes can pair up and move off slightly into
the bulk, the D3-brane can be considered more tightly bound to the
O3$^-$-plane than its companions. Eventually, the coupling is strong
enough so that the compound made of one D3-brane and an O3$^-$-plane is
better described as an O3$^+$-plane. Of course, this picture is rather
heuristic, but gives answers consistent with all constraints in the
system.
 
\subsection{Orientifold four-planes}

Supersymmetric configurations of D4-branes and O4-planes have been studied
in \cite{hori}, where their M-theory interpretation is provided (see also
\cite{gimon}). There are four kinds of configurations, which correspond to
$2P$ D4-branes on an O4$^-$-plane, $(2P+1)$ D4-branes on an O4$^-$-plane,
$2P$ D4-branes on and O4$^+$-plane or $2P$ D4-branes on an 
${\widetilde{O4}}^+$-plane. They differ in the choice of field-strength 
flux $\theta_{NS}$ for the NS-NS 2-form, and in the possibility of 
embedding the orientifold projection as a $\IZ_2$ Wilson line $w_R$ for
the RR $U(1)$ gauge field, as in \cite{schsen}. This information, the
charges under the RR 5-form, and the constraints of flux quantization in
M-theory \cite{witflux} are enough to provide the M-theory lifts of these
configurations, and therefore to study their strong coupling limits. The
result is as follows
{\small
\begin{center}
\begin{tabular}{|l|c|c|c|c|}
\hline
D-brane description   & $(\theta_{NS},w_R)$ & Charge & World-volume &
M-theory \\
\hline\hline
O4$^-$ + $2P$ D4     & $(0,0)$ & $2P-1$ & $SO(2P)$ & $\IR^5 \times 
\IR^5/\IZ_2 \times \IS^1$ + \\
 & & & & + $2P$ M5 \\
\hline
O4$^-$ + $(2P+1)$ D4  & $(0,\frac 12)$ & $2P$ & $SO(2P+1)$ & $\IR^5 \times
(\IR^5 \times \IS^1)/\IZ_2$ + \\
&&&& + $2P$ M5 \\
\hline
O4$^+$ + $2P$ D4     & $(\frac 12,0)$ & $2P+1$ & $USp(2P)$ & $\IR^5
\times \IR^5/\IZ_2 \times \IS^1$ + \\
&&&& + $(2P+2)$ M5 \\
\hline 
${\widetilde{O4}}^+$ + $2P$ D4 & $(\frac 12,\frac 12)$ & $2P+1$ &
$USp(2P)$ & $\IR^5 \times (\IR^5 \times \IS^1)/\IZ_2$ + \\
&&&& + $(2P+1)$ M5\\
\hline
\end{tabular}
\end{center}
}
The $\IZ_2$ acts by reflection of the coordinates of $\IR^5$, and, in the
second and fourth cases, by a half shift in the $\IS^1$ coordinate. The
M5-branes sit at the origin in the $\IR_5$ modded out by $\IZ_2$. In
matching the M5-brane charges with the D4-brane charges, one should take
into account that fixed points $\IR^5/\IZ_2$ in M-theory carry $-1$
fivebrane charge \cite{witorb}, while smooth geometries carry no fivebrane
charge. Also, one of the moduli in the M-theory configuration in the third
line is frozen \cite{hori}, by the mechanism explained in \cite{gimon}.

\medskip

We can repeat this exercise for the non-supersymmetric configurations of
$\ov{D4}$-branes on O4-planes. As in the previous section, the
classification in the supersymmetric case can be carried out for our
non-supersymmetric models. We can also propose suitable M-theory
configuration which reduce to these type IIA models, and are consistent
with flux quantization and other known properties of M-theory. It is
meaningful to consider such M-theory lifts because away from the
non-supersymmetric orientifold plane local physics is given by type IIA
physics. The following table should then be understood as providing the
appropriate M-theory objects to be placed in the orientifold core region
in the corresponding lifts
{\small
\begin{center}
\begin{tabular}{|l|c|c|c|c|}
\hline
D-brane description   & $(\theta_{NS},w_R)$ & Charge & World-volume &
M-theory \\
\hline\hline
O4$^-$ + $2P$ $\ov{D4}$ & $(0,0)$ & $-2P-1$ & $SO(2P)$ & $\IR^5 \times 
\IR^5/\IZ_2 \times \IS^1$ + \\
&&&& + $2P$ $\ov{M5}$ \\
\hline
O4$^-$ + $(2P+1)$ $\ov{D4}$  & $(0,\frac 12)$ & $-2P-2$ & $SO(2P+1)$ &
$\IR^5 \times (\IR^5 \times \IS^1)/\IZ_2$ + \\
&&&& $2P+2$ $\ov{M5}$ \\
\hline
O4$^+$ + $2P$ $\ov{D4}$ & $(\frac 12,0)$ & $-2P+1$ & $USp(2P)$ & $\IR^5
\times \IR^5/\IZ_2 \times \IS^1$ + \\
&&&& + $(2P-2)$ $\ov{M5}$ \\
\hline
${\widetilde{O4}}^+$ + $2P$ $\ov{D4}$ & $(\frac 12,\frac 12)$ & $2P+1$ &
$USp(2P)$ & $\IR^5 \times (\IR^5 \times \IS^1)/\IZ_2$ + \\
&&&& + $(2P-1)$ $\ov{M5}$\\
\hline
\end{tabular}
\end{center}
}
This information provides the strong coupling description of the
configurations of O4-planes and $\ov{D4}$-branes. However and as usual,
the most meaningful statements are restricted to the stable systems, and
in the following we center on the case of an O4$^-$-plane with one stuck
$\ov{D4}$-brane. In the strong coupling limit, this configuration is
better described in M-theory, as two $\ov{M5}$-branes in the background
geometry $\IR^5\times (\IR^5\times \IS^1)/\IZ_2$. Notice that the naive
lift as one $\ov{M5}$-brane in the background $\IR^5\times \IR^5/\IZ_2\times 
\IS^1$ is not consistent with the presence of non-zero $w_R$ in the
IIA model or with flux quantization in M-theory.

Notice that the two $\ov{M5}$-brane in M-theory are presumably bound to
the origin in $\IR^5$ due to attractive interactions. These cannot be
computed at short distances, given our ignorance about the fundamental
degrees of freedom in M-theory, but at long distances they reduce to the
exchange of massless supergravity fields. 

Another interesting observation involving short-distance physics in 
M-theory is that a phase transition seems to occur between the large and
small radius limits. Starting at large radius, and trying to  reach the
weakly coupled type IIA limit, beyond a certain radius the geometry $\IR^5
\times (\IR^5\times \IS^1)/\IZ_2$ is better described as a type IIA
O4$^-$-plane with an stuck D4-brane. The latter can thus annihilate with 
one of the antibranes present from the beginning, leading to our familiar
system of an O4$^-$-plane with one $\ov{D4}$-brane. This phase transition 
may imply that the M-theory lifts of these configurations are not useful in 
obtaining even qualitative features about gauge theories using brane
configurations as in \cite{hw}. 

\subsection{T-duality relations}

In this section we would like to use T-duality to relate our proposals for
the strong coupling behaviour of non-supersymmetric O3- and O4-planes, 
with an analysis inspired in \cite{gimon}. Given the equivalence between 
type IIB theory on a circle and M-theory on a 2-torus, one can find
strong-weak coupling duals in type IIB theory by obtaining two
different degenerations of the M-theory 2-torus. We illustrate this
technique in our non-supersymmetric orientifold context by considering
type IIB on $\IR^4\times \IR^5\times \IS^1$ modded out by $\Omega
(-1)^{F_L} I$, where $F_L$ is the left-handed world-sheet fermion number,
and $I$ inverts all coordinates of $\IR^5\times \IS^1$. The model contains
two O3-planes, which can be chosen of different type, and whose strong
coupling behaviour can now be derived from the M-theory realization. In
the following we consider several examples, with one O3-plane of type
$O3^-+{\ov{D3}}$ and one supersymmetric O3-plane.

\medskip

{\bf i)} Consider an initial configuration with an $O3^- + {\ov{D3}}$
system and an O3$^-$-plane, with a transverse circle $\IS^1$. Its M-theory
lift can be obtained by first T-dualizing to a type IIA model, and then
growing the M-theory circle ${\tilde{\IS}}^1$. In this case, the IIA
model is an $O4^-+{\ov{D4}}$ wrapped on $\IS^1$, and the M-theory lift
corresponds to two $\ov{M5}$-branes in the geometry $\IR^4\times \IS^1
\times (\IR^5\times {\tilde{\IS}}^1)/\IZ_2$. A different type IIB
description, corresponding to the strong coupling limit of the initial
one, can be now achieved by shrinking $\IS^1$ first and then T-dualizing
along ${\tilde{\IS}}^1$. Shrinking $\IS^1$ yields two $\ov{D4}$-branes in
the geometry $\IR^4\times (\IR^5\times {\tilde{\IS}}^1)/\IZ_2$, with a
$\IZ_2$ action not embedded as a $U(1)_R$ Wilson line ($w_R=0$). The T-dual 
of this configuration is given by two $\ov{D3}$-branes and two oppositely
charged O3-planes (see \cite{witseven}), whose overall $\theta_R$ must be
zero to agree with the vanishing type IIA $w_R$. The T-dual configuration
hence contains an $O3^++ 2{\ov{D3}}$ system and an O3$^-$-plane, which
precisely is the proposed strong coupling limit for the initial
configuration. Notice that the location of the $\ov{D3}$-branes on top of
the O3$^+$-plane obeys dynamical reasons.

{\bf ii)} Let us start with an $O3^-+{\ov{D3}}$ system and an
O3$^+$-plane. The T-dual configuration corresponds to one $\ov{D4}$-brane
in the geometry $\IR^4\times (\IR^5\times \IS^1)/\IZ_2$, with $w_R=1$. Its
M-theory lift is therefore one ${\ov{M5}}$-brane in the background
geometry $\IR^4 \times (\IR^5\times \IS^1\times {\tilde{\IS}^1})/\IZ_2$,
with the $\IZ_2$ acting with a simultaneous half-shift on both circles.
Notice this model is invariant under exchange of both circles, hence
shriking $\IS^1$ and T-dualizing along ${\tilde{\IS}^1}$ takes us to a type
IIB model isomorphic to the initial one. This self-duality is also obtained 
from our type IIB analysis, but in a non-trivial fashion. The strong 
coupling of the initial configuration is given, according to section 3.1,
by an $O3^++2{\ov{D3}}$ system and an $O3^-+D3$ system. The two 
$\ov{D3}$-branes are attracted by the $O3^+$, but more strongly by the
$D3$-brane (stuck at the O3$^-$-plane). Hence the true vacuum is achieved
only after annihilating a $D3$-${\ov{D3}}$ pair, and corresponds to an
O3$^+$-plane and an $O3^-+{\ov{D3}}$ system, as derived from the M-theory
argument. Notice that the agreement in this case is surprisingly intricate.

{\bf iii)} Consider an initial configuration of an $O3^-+{\ov{D3}}$ system 
and one ${\widetilde{O3}}^+$-plane. Its T-dual is given by one 
$\ov{D4}$-brane in the geometry $\IR^4\times (\IR^5\times\IS^1)/\IZ_2$,
and with $w_R=0$. Its M-theory lift is given by one ${\ov {M5}}$-brane in
the geometry $\IR^4\times {\tilde{\IS}}^1\times(\IR^5\times \IS^1)/\IZ_2$.
Upon shrinking $\IS^1$, we recover an ${\widetilde{O4}}^+ + 2{\ov{D4}}$
system (this is more easily understood by lifting the IIA configuration,
and annihilating a $M5$-$\ov{M5}$ pair). Finally, T-dualizing to type IIB,
we recover an $O3^++2{\ov{D3}}$ system and an ${\widetilde{O3}}^+$-plane,
which agrees with our strong coupling proposal of the initial configuration.

{\bf iv)} Finally, consider an $O3^-+{\ov{D3}}$ system and an $O3^-+D3$
system. The T-dual IIA model corresponds to one $O4^-$-plane with one
stuck D4-brane and one stuck ${\ov{D4}}$-brane, with different Wilson
lines along the $\IS^1$ they wrap, and with $w_R=0$. In M-theory this is
described as one M5-brane and one ${\ov{M5}}$-brane on $\IR^4\times
\IS^1\times \IR^5/\IZ_2\times {\tilde{\IS}}^1$, with different `Wilson
lines' (actually, periods of the world-volume self-dual 2-form on the
M-theory 2-torus). After shrinking $\IS^1$, we obtain a IIA configuration 
of an $O4^++2{\ov{D4}}$ system , with no Wilson lines (again this is
easier to understand by lifting the IIA configuration, and annihilating
a $M5$-$\ov{M5}$ pair with identical `Wilson lines'). The type IIB T-dual
contains an $O3^++2{\ov{D3}}$ system and an O3$^+$-plane, which agrees
with the strong coupling proposal for the initial model.

\medskip

Notice that the above arguments involve shriking circles in M-theory,
whose treatment is not completely rigorous in the absence of supersymmetry
and so of the BPS property. Therefore they rely in the assumption that
supersymmetry away from the orientifold core is enough to allow taking
such limits. Notice also that even though the duality chains are quite
constrained from mere `kinematics', namely matching of charges, fluxes,
etc (that is actually the reason that allows us to match non-supersymmetric 
configurations) there is some role played by non-trivial dynamics, in
particular in the form of brane-antibrane annihilations, and of uncancelled 
antibrane-orientifold forces. Finally notice that in the above discussion
we have ignored the issue of the dynamics of the modulus associated to the
circle radius, which in a more detailed treatment should perhaps also be
taken into account.

The above examples mainly center on our proposal for the strong coupling
limit of the $O3^-+{\ov{D3}}$ system. Other examples can be studied
analogously.

\subsection{Orientifold two-planes}

We conclude this Section with a brief discussion on the strong coupling
description of an $O2^-$-plane with one stuck $\ov{D2}$-brane. The only
ingredients required from the supersymmetric case are the M-theory lifts 
of the O2$^-$-plane, the O2$^+$-plane and the $O2^-+{\ov{D2}}$ system,
determined in \cite{sethi}. They correspond to M-theory geometries of the
form $\IR^3\times (\IR^7\times \IS^1)/\IZ_2$, where the $\IZ_2$ action
reverses all the coordinates of $\IR^7\times \IS^1$. The model contains
two fixed points, locally of the form, $\IR^8/\IZ_2$, and each having
one of two possible values of field-strength flux for the M-theory 3-form. 
The two possibilites endow the fixed point with different membrane charges: 
$-1/8$ for a singularity with vanishing flux and $3/8$ for a singularity
with non-zero flux. The M-theory descriptions of the relevant O2-planes are
{\small
\begin{center}
\begin{tabular}{|l|c|c|}
\hline
IIA description & Charge & M-theory fixed points \\
\hline\hline
O2$^-$ & $-1/4$ & $(-1/8,-1/8)$ \\
\hline
O2$^-$ + $\ov{D2}$ & $3/4$ & (3/8,3/8) \\
\hline
O2$^+$ & $1/4$ & $(3/8,-1/8)$ \\
\hline
\end{tabular}
\end{center}
}
In our context of non-supersymmetric configurations of O2-planes and
$\ov{D2}$-branes, we center on the particular case of the $O2^-+{\ov{D2}}$
system (other examples can be worked out analogously). Using our
experience in similar systems in other dimensions, we propose the correct
M-theory lift is given by $\IR^3\times(\IR^7\times \IS^1)/\IZ_2$ with two
fixed points of charge $3/8$ and two $\ov{M2}$-branes. 

In order to show that, we use T-duality with some familiar configurations
of O3-planes and $\ov{D3}$-branes. Consider an $O3^-+{\ov{D3}}$ system,
wrapped on a longitudinal $\IS^1$, and perform a T-duality along it. The
resulting type IIA model contains an O2$^-$-plane and an $O2^-+{\ov{D2}}$
system, with a transverse circle. The M-theory lift of this configuration
is given by $\IR^3\times (\IR^6\times \IT^2)/\IZ_2$, with two
$\ov{M2}$-branes. There are four fixed points, two of which have charge
$-1/8$ (from lifting the O2$^-$-plane) and two have charge $3/8$ (from our
proposed lift of the $O2^-+{\ov{D2}}$ system). This M-theory lift can be
confirmed by first S-dualizing the initial type IIB model, and then
lifting it to M-theory. In the strong coupling limit, the initial IIB
configuration turns into an $O3^+$ with two $\ov{D3}$-branes, wrapped on
$\IS^1$. Its type IIA T-dual contains two O2$^+$-planes and two
$\ov{D2}$-branes. Its M-theory lift corresponds to $\IR^3\times(\IR^6
\times \IT^2)/\IZ_2$, with two $\ov{M2}$-branes, two fixed points of
charge $-1/8$ and two of charge $3/8$ (since each O2$^+$ contributes with
one fixed point of each kind). This agrees with the M-theory configuration
found before, and supports our identification of the strong coupling limit
of the $O2^-+{\ov{D2}}$ system.

Notice that the naive M-theory lift corresponding to M-theory on $\IR^3 
\times (\IR^7\times \IS^1)/\IZ_2$ with two fixed points of charge $-1/8$,
and with one $\ov{M2}$-brane is not correct. In fact, having M2-branes (or
$\ov{M2}$-branes) stuck at $\IR^8/\IZ_2$ singularities is not consistent
with flux quantization conditions.

We hope this example suffices to illustrate the discussion of O2-planes,
and spare the reader an exhaustive treatment, already performed in the
analogous case of O3- and O4-planes.

\bigskip

\section{Final remarks}

The purpose of this paper has been to explore some of the properties of
systems of $\ov{Dp}$-branes on O$p$-planes of different kinds. We believe
these configurations show interesting features and non-supersymmetric
dynamics, which nevertheless seem accessible to study (to a certain
extent) due to the simplicity of the configuration.

A particular avenue, not directly exploited in this note, is to consider 
the limit of a large number $N$ of $\ov{Dp}$-branes on the orientifold
planes \footnote{I thank G.~Mandal for conversations on this point.}. This
approach would be particularly useful to study the field theories in the
$\ov{Dp}$-brane world-volume, by computing in the dual supergravity
background in the sense of the AdS/CFT correspondence (see \cite{ads} for a 
review). Some results in this direction have appeared in \cite{berkooz}
for the case of $\ov{M2}$-branes. For reasons already explained, the
bosonic parts of the corresponding supergravity backgrounds are identical
to those in the supersymmetric cases. For instance, for $\ov{D3}$-brane on
O3-planes, the background at leading order in $N$ is given by $\AdS_5\times 
\RP_5$, just as in the supersymmetric case in \cite{witthree}. The
supersymmetry breaking effects arise because fermionic fields pick up an
additional $(-1)$ in going along non-contractible cycles in $\RP_5$, as
compared with the supersymmetry-preserving case \footnote{This is
equivalent to the orientation-reversing introduced in \cite{duff} in
more general supergravity backgrounds.}. This effect is subleading in $N$,
since it arises from the orientifold projection, which is suppressed in the
large $N$ limit \cite{bkv}. It would be interesting to study $1/N$
corrections to different quantities in these type of backgrounds (for
instance, the subleading correction to the conformal anomaly, analyzed in 
\cite{narain} in the supersymmetric case (see also \cite{odintsov} for
related computations)).

A different line of development would be to consider other values of $p$.
In some cases this would require improving our understanding of the
different orientifold planes even in the supersymmetric case. Finally,
it would be interesting to consider more complicated configurations, by
introducing new objects ({\em e.g.} additional branes and/or antibranes)
or orbifold projections. This would require further developments in the
study of the stability of additional moduli in the former case, and the
appropriate treatment of uncancelled twisted NS tadpoles in the latter.

In any event, we hope our observations on these systems are useful as a
starting point for the investigation on these models.

\bigskip

\centerline{\bf Acknowledgements}

I am pleased to thank G.~Aldazabal, J.~L.~F.~Barb\'on, L.~E.~Ib\'a\~nez
and R.~Rabad\'an for useful discussions, and M.~Gonz\'alez for
encouragement and support.


\begin{thebibliography}{99}
%
\bibitem{orient}
A.~Sagnotti in {\em Cargese' 87} ``Non-perturbative Quantum Field
Theory'', ed. G.~Mack et al. (Pergamon Press 88), pag. 521; ``Some
properties of open string theories'', hep-th/9509080. \\
%
J.~Dai, R.~G.~Leigh, J.~Polchinski, ``New connections between string
theories'', \MODA{4}{89}{2073}; R.~G.~Leigh, ``Dirac-Born-Infeld Action
From Dirichlet Sigma Model'', \MODA{4}{89}{2767}. \\
%
P.~Horava, ``Strings on world-sheet orbifolds'', \NPB{327}{89}{461};
``Background duality of open string models'', \PLB{231}{89}{251};
``Two-dimensional stringy black holes with one asymptotically flat
domain'', \PLB{289}{92}{293}; ``Equivariant topological sigma models'',
\NPB{418}{94}{571}. \\
%
G.~Pradisi, A.~Sagnotti, ``Open strings orbifolds'', \PLB{216}{89}{59};
%
M.~Bianchi, A.~Sagnotti, ``On the systematics of open string theories'',
\PLB{247}{90}{517}; ``Twist symmetry and open string Wilson lines'',
\NPB{361}{91}{519}.
%
\bibitem{tasi}
J.~Polchinski, S.~Chaudhuri, C.~V.~Johnson, ``Notes on D-branes'', 
hep-th/9602052; J.~Polchinski, ``Tasi lectures on D-branes'',
hep-th/9611050.
%
\bibitem{nonsusy}
I.~Antoniadis, E.~Dudas, A.~Sagnotti, ``Brane Supersymmetry Breaking'',
\PLB{464}{99}{38}, hep-th/9908023. \\
%
G.~Aldazabal, A.~M.~Uranga, ``Tachyon-free Non-supersymmetric Type IIB
Orientifolds via Brane-Antibrane Systems'', JHEP 9910 (1999) 024,
hep-th/9908072. \\
%
G.~Aldazabal, L.~E.~Ib\'a\~nez, F.~Quevedo, ``Standard-like Models with
Broken Supersymmetry from Type I String Vacua'', hep-th/9909172. \\
%
C.~Angelantonj, I.~Antoniadis, G.~D'Appollonio, E.~Dudas, A.~Sagnotti,
``Type I vacua with brane supersymmetry breaking'', hep-th/9911081;
I.~Antoniadis, A.~Sagnotti, ``Mass scales, supersymmetry breaking and open
strings'', hep-th/9911205.
%
\bibitem{sugimoto}
S.~Sugimoto, ``Anomaly cancellations in type I D9 - anti-D9 system and
the USp(32) string theory'',  Prog.Theor.Phys. 102(1999)685,
hep-th/9905159. 
%
\bibitem{banks}
T.~Banks, L.~Susskind, ``Brane-antibrane forces'', hep-th/9511194.
%
\bibitem{witthree}
E.~Witten, ``Baryons and branes in anti-de Sitter space'', JHEP
9807(1998)006, hep-th/9805112.
%
\bibitem{egkt}
S.~Elitzur, A.~Giveon, D.~Kutasov, D.~Tsabar, ``Branes, orientifolds and
chiral gauge theories'', \NPB{524}{98}{251}, hep-th/9801020.
%
\bibitem{hori}
K.~Hori, ``Consistency condition for five-brane in M theory on
$\IR^5/\IZ_2$ orbifold'', \NPB{539}{99}35, hep-th/9805141.
%
\bibitem{gimon}
E.~G.~Gimon, ``On the M theory interpretation of orientifold
planes'', hep-th/9806226.
%
\bibitem{sethi}
S.~Sethi, ``A Relation between N=8 gauge theories in three-dimensions'', 
JHEP 9811(1998)003, hep-th/9809162.
%
\bibitem{hanany}
A.~Hanany, B.~Kol, A.~Rajaraman, ``Orientifold points in M theory'',
JHEP 9910(1999)027, hep-th/9909028. 
%
\bibitem{witfive}
E.~Witten, ``New 'gauge' theories in six-dimensions'', JHEP 9801(1998)001, 
Adv.Theor.Math.Phys. 2(1998)61, hep-th/9710065.
%
\bibitem{landlop}
K.~Landsteiner, E.~Lopez, ``New curves from branes'', \NPB{516}{98}{273},
hep-th/9708118. 
%
\bibitem{witseven}
E.~Witten, ``Toroidal compactification without vector structure'', JHEP
9802(1998)006, hep-th/9712028.
%
\bibitem{sen}
A.~Sen, ``Duality and orbifolds'', \NPB{474}{96}{361}, hep-th/9604070.
%
\bibitem{thooft}
G.~'t~Hooft, in ``Recent Developments in Gauge Theories'', eds.
G.~'t~Hooft {\em et al.} (Plenum Press, New York, 1980).
%
\bibitem{schsen}
J.~H.~Schwarz, A.~Sen, ``Type IIA dual of the six-dimensional CHL
compactification'', \PLB{357}{95}{323}, hep-th/9507027.
%
\bibitem{witflux}
E.~Witten, ``On Flux Quantization In M-Theory And The Effective Action'',
J.Geom.Phys. 22(1997)1, hep-th/9609122.
%
\bibitem{witorb}
E.~Witten, ``Fivebranes and M-theory on an orbifold'', \NPB{463}{96}{383},
hep-th/9512219 
%
\bibitem{hw}
A.~Hanany, E.~Witten, ``Type IIB superstrings, BPS monopoles, and
three-dimensional gauge dynamics'', \NPB{492}{97}{152}, hep-th/9611230;
E.~Witten, ``Solutions of four-dimensional field theories via M theory'',
\NPB{500}{97}{3}, hep-th/9703166; ``Branes and the dynamics of QCD'',
\NPB{507}{97}{658}, hep-th/9706109.
%
\bibitem{ads}
O.~Aharony, S.~S.~Gubser, J.~Maldacena, H.~Ooguri, Y.~Oz, ``Large N field
theories, string theory and gravity'', hep-th/9905111.
%
\bibitem{berkooz}
M.~Berkooz, S.-J.~Rey, ``Nonsupersymmetric stable vacua of M theory'', 
JHEP 9901(1999)014, hep-th/9807200; M.~Berkooz, A.~Kapustin, ``A Comment
on nonsupersymmetric fixed points and duality at large N'',
hep-th/9903195.
%
\bibitem{duff}
M.~J.~Duff, B.~E.~W.~Nilsson, C.~N.~Pope, ``Spontaneous supersymmetry
breaking by the squashed seven-sphere'', \PRL{50}{83}{2043}, {\em Err.
ib.} 51(1983)846; ``Kaluza-Klein supergravity'', \PRT{130}{86}{1}.
%
\bibitem{bkv}
M.~Bershadsky, Z.~Kakushadze, C.~Vafa, ``String expansion as large N
expansion of gauge theories'', \NPB{523}{98}{59}, hep-th/9803076; 
Z.~Kakushadze, ``Gauge theories from orientifolds and large N limit'',
\NPB{529}{98}{157}, hep-th/9803214.
%
\bibitem{narain}
M.~Blau, E.~Gava, K.~S.~Narain, ``On Subleading Contributions to the
AdS/CFT Trace Anomaly'', JHEP 9909(1999)018, hep-th/9904179.
%
\bibitem{odintsov}
S.~Nojiri, S.~D.~Odintsov, ``On the conformal anomaly from higher
derivative gravity in AdS/CFT correspondence'', hep-th/9903033; ``Weyl
anomaly from Weyl gravity'', hep-th/9910113. 
%
\end{thebibliography}
\end{document}